\documentclass[a4paper,journal,11pt,draftclsnofoot,onecolumn]{IEEEtran}

\usepackage{dsfont}
\usepackage{amsmath}
\usepackage{amssymb}
\usepackage{amsfonts}
\usepackage{graphicx,dblfloatfix}
\usepackage{amsmath}
\usepackage[noadjust]{cite}
\usepackage[all,poly]{xy}
\usepackage{multirow,balance}
\usepackage{rotate}

\def\bfa{{\mathbf{a}}}

\def\bfg{{\mathbf{g}}}

\def\bfm{{\mathbf{m}}}
\def\bfn{{\mathbf{n}}}

\def\bfw{{\mathbf{w}}}
\def\bfx{{\mathbf{x}}}
\def\bfy{{\mathbf{y}}}

\def\bfM{{\mathbf{M}}}

\def\calH{{\mathcal{H}}}

\def\calR{{\mathcal{R}}}

\newcommand{\figwidth}{0.90\columnwidth}
\newcommand{\figwidththird}{0.3\textwidth}

\title{Nonlinear Unmixing of Hyperspectral Images: Models and Algorithms}

\author{\vspace{1cm}Nicolas Dobigeon$^{(1)}$, Jean-Yves
Tourneret$^{(1)}$, C\'edric Richard$^{(2)}$, \\Jos\'e C. M.
Bermudez$^{(3)}$, Stephen
McLaughlin$^{(4)}$, and Alfred O. Hero$^{(5)}$\\
\vspace{1cm}
\normalsize $^{(1)}$ University of Toulouse, IRIT/INP-ENSEEIHT/T\'eSA, Toulouse, France\\
\small\texttt{\{Nicolas.Dobigeon,Jean-Yves.Tourneret\}@enseeiht.fr}\\
\normalsize $^{(2)}$ Universit\'e de Nice Sophia-Antipolis, OCA, Laboratoire Lagrange, Nice Cedex 2, France\\
\small\texttt{cedric.richard@unice.fr}\\
\normalsize $^{(3)}$ University of Santa Catarina, Florian\'opolis, SC, Brazil\\
\small\texttt{j.bermudez@ieee.org}\\
\normalsize $^{(4)}$ School of Engineering and Physical Sciences, Heriot-Watt University, Edinburgh, U.K.\\
\small\texttt{s.mclaughlin@hw.ac.uk}\\
\normalsize $^{(5)}$ University of Michigan, Department of EECS, Ann Arbor, USA.\\
\small\texttt{hero@umich.edu}}

\begin{document}
 \maketitle

%% ABSTRACT
\begin{abstract}
  When considering the problem of unmixing hyperspectral images, most
of the literature in the geoscience and image processing areas
relies on the widely used linear mixing model (LMM). However, the
LMM may be not valid and other nonlinear models need to be
considered, for instance, when there are multi-scattering effects or
intimate interactions. Consequently, over the last few years,
several significant contributions have been proposed to overcome the
limitations inherent in the LMM. In this paper, we present an
overview of recent advances in nonlinear unmixing modeling.

% The main
%nonlinear models are introduced and their validity discussed. Then,
%we describe the main classes of unmixing strategies designed to
%solve the problem in supervised and unsupervised frameworks.
%Finally, the problem of detecting nonlinear mixtures in
%hyperspectral images is addressed.
%Finally, several open
%challenges conclude the paper.

\end{abstract}

\newpage
%%% INTRODUCTION
\section{\small Motivation for nonlinear models}
\label{sec:introduction}

Spectral unmixing (SU) is widely used for analyzing hyperspectral
data arising in areas such as: remote sensing, planetary science
chemometrics,  materials science and other areas of
micro-spectroscopy. SU provides a comprehensive and quantitative
mapping of the elementary materials that are present in the acquired
data. More precisely, SU can identify the spectral signatures of
these materials (usually called \emph{endmembers}) and can estimate
their relative contributions (or \emph{abundances}) to the measured
spectra. Similar to other blind source separation tasks, the SU
problem is naturally ill-posed and admits a wide range of admissible
solutions. As a consequence, SU is a challenging problem that has
received considerable attention in the remote sensing, signal and
image processing communities \cite{Bioucas2012jstars}. Hyperspectral
data analysis can be \emph{supervised}, when the endmembers are
known, or \emph{unsupervised}, when they are unknown. Irrespective
of the case, most SU approaches require the definition of the mixing
model underlying the observations. A mixing model describes in an
analytical fashion how the endmembers combine to form the mixed
spectrum measured by the sensor. The abundances parametrize the
model. Given the mixing model, SU boils down to estimating the
inverse of this formation process to infer the quantities of
interest, namely the endmembers and/or the abundances, from the
collected spectra. Unfortunately, defining the direct observation
model that links these meaningful quantities to the measured data is
a non-trivial issue, and requires a thorough understanding of
complex physical phenomena. A model based on radiative transfer (RT)
could accurately describe the light scattering by the materials in
the observed scene \cite{Hapke1981}, but would lead to very complex
unmixing problems. Fortunately, invoking simplifying assumptions can
lead to exploitable mixing models.

When the mixing scale is macroscopic and each photon reaching the
sensor has interacted with just one material, the measured spectrum
$\bfy_p \in\mathbb{R}^L$ in the $p$th pixel can be accurately
described by the linear mixing model
\begin{equation}
\label{eq:LMM}
  \bfy_p = \sum_{r=1}^R a_{r,p}\bfm_r + \bfn_p
\end{equation}
where $L$ is the number of spectral bands, $R$ is the number of
endmembers present in the image, $\bfm_r$ is the spectral signatures
of the $r$th endmember, $a_{r,p}$ is the abundance of the $r$th
material in the $p$th pixel and $\bfn_p$ is an additive term
associated with the measurement noise and the modeling error. The
abundances can be interpreted as the relative areas occupied by the
materials in a given image pixel~\cite{Hapke1993book}. Thus it is
natural to consider additional constraints regarding the abundance
coefficients $a_{r,p}$
\begin{equation}
\label{eq:LMM_constraints}
  \left\{
     \begin{array}{ll}
       a_{r,p}\geq 0, & \forall p,\ \forall r \\
       \sum_{r=1}^R a_{r,p}=1, & \forall p
     \end{array}
   \right.
\end{equation}
In that case, SU can be formulated as a constrained blind source
separation problem, or constrained linear regression, depending on
the prior knowledge available regarding the endmember spectra.

\begin{figure}[h!]
  \vspace{-0.15cm}
  \centering
  \includegraphics[width=\figwidth]{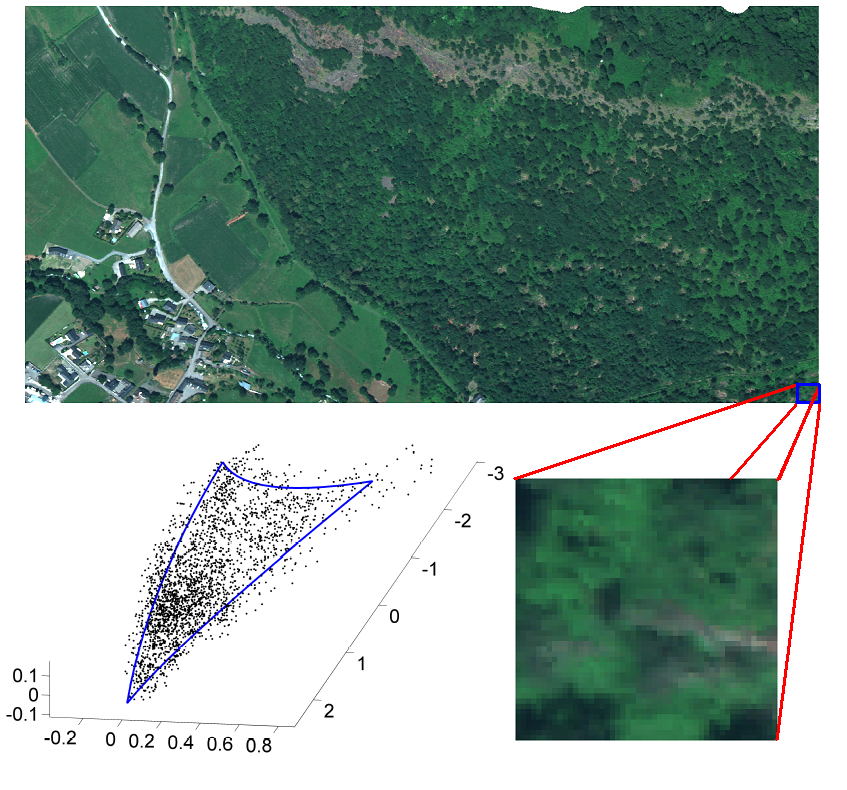}
  \vspace{-0.3cm}
  \caption{Top: real hyperspectral Madonna data acquired by the Hyspex
hyperspectral scanner over Villelongue, France. Bottom, right: the
region of interest shown in true colors. Bottom, left:
Representation of the $N = 2500$ pixels (black dots) of the data and
boundaries of the estimated nonlinear simplex (blue lines).}
\label{fig:Madonna}
  \vspace{-0.15cm}
\end{figure}

Due to the relative simplicity of the model and the straightforward
interpretation of the analysis results, LMM-based unmixing
strategies predominate in the literature. All of these techniques
have been shown to be very useful whenever the LMM represents a good
approximation to the actual mixing. There are, however, practical
situations in which the LMM is not a suitable approximation
\cite{Bioucas2012jstars}. As an illustrative example, consider a
real hyperspectral image, composed of $L=160$ spectral bands from
the visible to near infrared, acquired in 2010 by the airborne
Hyspex hyperspectral sensor over Villelongue, France. This image,
with a spatial resolution of $0.5$m, is represented in Fig.
\ref{fig:Madonna} (top). From primary inspection and prior knowledge
coming from available ground truth, the $50\times 50$ pixel region
of interest depicted in Fig. \ref{fig:Madonna} (bottom, right) is
known to be composed of mainly $R=3$ macroscopic components (oak
tree, chestnut tree and an additional non-planted-tree component).
When considering the LMM to model the interactions between these
$R=3$ components, all the observed pixels should lie in a
$2$-dimensional linear subspace, that can be easily identified by a
standard principal component analysis (PCA). Conversely, if
nonlinear effects are present in the considered scene, the observed
data may belong to a $2$-dimensional nonlinear manifold. In that
case, more complex nonlinear dimension reduction procedures need to
be considered. The accuracy of these dimension reduction procedures
in representing the dataset into a $2$-dimensional linear or
nonlinear subspace can be evaluated thanks to the average
reconstruction error (ARE), defined as
\begin{equation}
  \mathrm{ARE} = \sqrt{\frac{1}{LP}\sum_{n=1}^N \left\|\bfy_n-\hat{\bfy}_{n}\right\|^2}
\end{equation}
where $\bfy_n$ are the observed pixels and $\hat{\bfy}_{n}$ the
corresponding estimates.  Here we contrast two approaches, a locally
linear Gaussian process latent variable model (LL-GPLVM) introduced
in \cite{Altmann2013sp} and PCA. When using PCA to represent the
data, the obtained ARE is $8.4\times10^{-3}$ while using the
LL-GPLVM, the ARE is reduced to $7.9\times10^{-3}$. This
demonstrates that the investigated dataset should be preferably
represented in a nonlinear subspace, as clearly demonstrated in Fig.
\ref{fig:Madonna} (bottom, left), where the nonlinear simplex
identified by the fully constrained LL-GPLVM has been represented as
blue lines.

Consequently, more complex mixing models need to be considered to
cope with nonlinear interactions. These models are expected to
capture important nonlinear effects that are inherent
characteristics of hyperspectral images in several applications.
They have proven essential to unveil meaningful information for the
geoscience community
\cite{Ray1996,Combe2005,Liu2005,Arai2008,Fan2009,Somers2009}.
Several approximations to the RT theory have been proposed, such as
Hapke's bidirectional model \cite{Hapke1993book}. Unfortunately,
these models require highly non-linear and integral formulations
that hinder practical implementations of unmixing techniques. To
overcome these difficulties, several physics-based approximations of
Hapke's model have been proposed, mainly in the spectroscopy
literature (e.g., see \cite{Hapke1993book}). However, despite their
wide interest, these approximations still remain difficult to apply
for automated hyperspectral imaging. In particular, for such models,
there is no unsupervised nonlinear unmixing algorithm able to
jointly extract the endmembers from the data and estimate their
relative proportions in the pixels. Meanwhile, several approximate
but exploitable nonlinear mixing models have been recently proposed
in the remote sensing and image processing literatures. Some of them
are similarly motivated by physical arguments, such as the class of
bilinear models introduced later. Others exploit a more flexible
non-linear mathematical model to improve unmixing performance.
Developing effective unmixing algorithms based on nonlinear mixing
models represents a challenge for the signal and image processing
community.  Supervised and unsupervised algorithms need to be
designed to cope with nonlinear transformations that can be
partially or totally unknown. Solving the nonlinear unmixing problem
requires innovative approaches to existing signal processing
techniques.\\

More than 10 years after Keshava and Mustard's  comprehensive review
article on spectral unmixing \cite{Keshava2002}, this article
provides an updated review focusing on non-linear unmixing
techniques introduced in the past decade. In \cite{Keshava2002}, the
problem on nonlinear mixtures was thoroughly addressed but, at that
time, very few algorithmic solutions were available. Capitalizing on
almost one decade of advances in solving the linear unmixing
problem, scientists from the signal and image processing communities
have developed, and continue to do so, automated tools to extract
endmembers from nonlinear mixtures, and to quantify their
proportions in nonlinearly mixed pixels. The paper is organized as
follows. The principal nonlinear mixing models are presented in the
next section. Then the most popular nonlinear unmixing algorithms
are reviewed. Model-based and model-free algorithms are considered.
Existing solutions for supervised and unsupervised unmixing are also
discussed. At the end of this paper, we present some recent
strategies for detection of nonlinear mixtures in hyperspectral
data. Finally, challenges and future directions for hyperspectral
unmixing are reported in the conclusions.

%%% MODELS
\section{\small Non linear models}
\label{sec:models}

\begin{figure*}[t!]
  \centering
\begin{minipage}[b]{\figwidththird}
  \includegraphics[width=\textwidth]{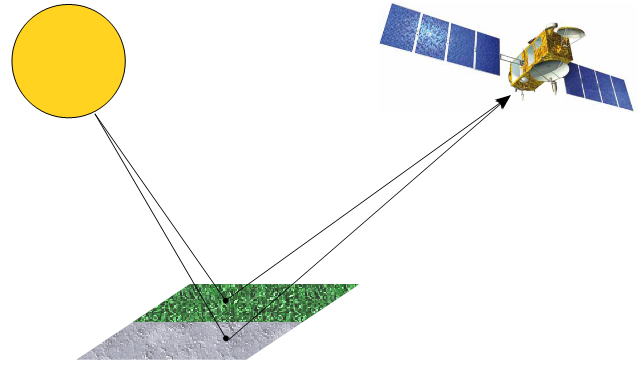}
  % \label{fig:LMM}
%  \vspace{2.0cm}
  \centerline{(a)}%\medskip
\end{minipage}
\hfill
\begin{minipage}[b]{\figwidththird}
  \includegraphics[width=\textwidth]{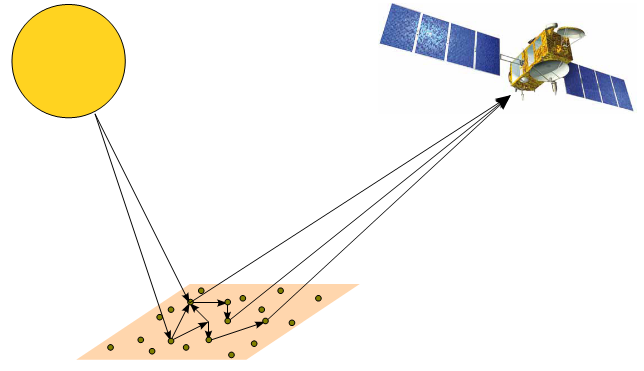}
  %\label{fig:IM}
%  \vspace{1.5cm}
  \centerline{(b)}%\medskip
\end{minipage}
\hfill
\begin{minipage}[b]{\figwidththird}
  \includegraphics[width=\textwidth]{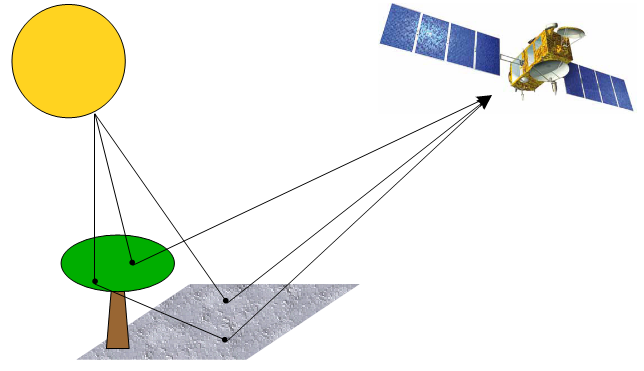}
  %\label{fig:BM}
%  \vspace{1.5cm}
  \centerline{(c)}%\medskip
\end{minipage}
\vspace{-0.3cm}
 \caption{(a) Linear mixing model: the imaged pixel
is composed of two materials. (b) Intimate mixture: the imaged pixel
is composed of a microscopic mixture of several constituents. (c)
Bilinear model: the imaged pixel is composed of two endmembers,
namely tree and soil. In addition to the individual contribution of
each material, bilinear interactions between the tree and the soil
reach the sensor.} \label{fig:all_models} \vspace{-0.3cm}
\end{figure*}

In \cite{Bioucas2012jstars}, it is explained that linear mixtures
are reasonable when two assumptions are wholly fulfilled. First the
mixing process must occur at a macroscopic scale \cite{Singer1979}.
Secondly, the photons that reach the sensor must interact with only
one material, as is the case in checkerboard type scenes
\cite{Clark1984}. An illustration of this model is depicted in Fig.
\ref{fig:all_models}(a) for a scene composed of two materials. When
one of these two assumptions does not hold, different nonlinear
effects may occur. Two families of nonlinear models are described in
what follows.

\subsection{Intimate mixtures}

The first assumption for linear mixtures is a macroscopic mixing
scale. However, there are common situations when interactions occur
at a microscopic level. The spatial scales involved are typically
smaller than the path length followed by the photons. The materials
are said to be intimately mixed \cite{Hapke1993book}. Such mixtures
have been observed and studied for some time, e.g., for imaged
scenes composed of sand or mineral mixtures \cite{Nash1974}. They
have been advocated for analyzing mixtures observed in laboratory
\cite{Mustard1989}. Based on RT theory, several theoretical
frameworks have been derived to accurately describe the interactions
suffered by the light when encountering a surface composed of
particles.

%\begin{figure}[h!]
%  \centering
%  \includegraphics[width=\figwidthbis]{fig_IM_new.png}
%  \caption{Intimate mixture: the imaged pixel is composed of a microscopic mixture of several constituents.} \label{fig:IM}
%\end{figure}

An illustration of these interactions is represented in Fig.
\ref{fig:all_models}(b). Probably the most popular approaches
dealing with intimate mixtures are those of Hapke in
\cite{Hapke1993book} since they involve meaningful and interpretable
quantities that have physical significance. Based on these concepts,
several simplified nonlinear mixing models have been proposed to
relate the measurements to some physical characteristics of the
endmembers and to their corresponding abundances (that are
associated with the relative mass fractions for intimate mixtures).
In \cite{Hapke1981}, the author derives an analytical model to
express the measured reflectances as a function of parameters
intrinsic to the mixtures, e.g., the mass fractions, the
characteristics of the individual particles (density, size) and the
single-scattering albedo. Other popular approximating models include
the discrete-dipole approximation \cite{Draine1988} and the
Shkuratov's model \cite{Shkuratov1999b} (interested readers are
invited to consult \cite{Hapke1993book} or the more signal
processing-oriented papers \cite{Nascimento2010,Close2012spie}).
However these models also strongly depend on parameters inherent to
the experiment since it requires the perfect knowledge of the
geometric positioning of the sensor with respect to the observed
sample. This dependency upon external parameters makes the inversion
(i.e., the estimation of the mass fractions from the collected
spectra) very difficult to implement and, obviously, even more
challenging in a unsupervised scenario, i.e., when the spectral
signatures of the materials are unknown and need to be also
recovered.

More generally, it is worth noting that the first requirement of
having a macroscopic mixing scale is intrinsically related to the
definition of the endmembers. Indeed, defining a pure material
requires specification of the spatial or spectral resolution, which
is application dependent. Consider, for example, a simple scene
composed of $3$ materials $A$, $B$ and $C$. It is natural to expect
retrieval of these components individually when analyzing the scene.
However, in other circumstances, one may be interested in the
material components themselves, for instance, $A_1$, $A_2$, $B_1$,
$B_2$, $C_1$ and $C_2$ if we assume that each material is composed
of $2$ constituents. In that case, pairs of subcomponents combine
and, by performing unmixing, one might also be interested in
recovering each of these $6$ components. Conversely, it may be well
known that the material $A$ can never be present in the observed
scene without the material $B$. In such case, unmixing would consist
of identifying the couple $A+B$ and $C$, without distinguishing the
subcomponent $A$ from the subcomponent $B$. This issue is frequently
encountered in automated spectral unmixing. In each scenario, it is
clear that when more details are desired, the mixtures should not
occur at a macroscopic scale. To circumvent this difficulty in
defining the mixture scale, it makes sense to associate pure
components with individual instances whose resolutions have the same
order of magnitude than the sensor resolution. For example, a patch
of sand of spatially homogeneous composition can be considered as a
unique pure component. In that case, most of the interactions
occurring in most of the scenes of interest can be reasonably
assumed to occur at a macroscopic level, at least when analyzing
airborne and spaceborne remotely sensed images.

\subsection{Bilinear models}

Another type of nonlinear interaction occurs at a macroscopic scale,
in particular in so-called \emph{multilayered} configurations.  One
may encounter this nonlinear model when the light scattered by a
given material reflects off other materials before reaching the
sensor. This is often the case for scenes acquired over forested
areas, where there may be many interactions between the ground and
the canopy. An archetypal example of this kind of scene is shown in
Fig. \ref{fig:all_models}(c).

%\begin{figure}[h!]
%  \centering
%  \includegraphics[width=\figwidthbis]{fig_BM_new.png}
%  \caption{Bilinear model: the imaged pixel is composed of two endmembers, namely tree and soil. In addition to the individual contribution of each material,
%bilinear interactions between the tree and the soil reach the
%sensor.} \label{fig:BM}
%\end{figure}

Several models have been proposed to analytically describe these
interactions. They consist of including powers of products of
reflectance. However they are usually employed such that
interactions of orders greater than two are neglected. The resulting
models are known as the family of the bilinear mixing models.
Mathematically, for most of these bilinear models, the observed
spectrum $\bfy_p \in \mathbb{R}^L$ in $L$ spectral bands for the
$i$th pixel is approximated by the following expansion
\begin{equation}
\label{eq:bilinear_model}
  \bfy_p = \sum_{r=1}^R a_{r,p}\bfm_r + \sum_{i=1}^{R-1}\sum_{j=i+1}^R \beta_{i,j,p} \bfm_i \odot \bfm_j +
  \bfn_p.
\end{equation}
where $\odot$ stands for the termwise (Hadamard) product
\begin{equation*}
  \bfm_i \odot \bfm_j = \left(
                          \begin{array}{c}
                            m_{1,i} \\
                            \ldots \\
                            m_{L,i} \\
                          \end{array}
                        \right) \odot
                        \left(
                          \begin{array}{c}
                            m_{1,j} \\
                            \ldots \\
                            m_{L,j} \\
                          \end{array}
                        \right) =
                        \left(
                          \begin{array}{c}
                            m_{1,i}m_{1,j} \\
                            \ldots \\
                            m_{L,i}m_{L,j} \\
                          \end{array}
                        \right)
\end{equation*}
In the right-hand side of \eqref{eq:bilinear_model}, the first term,
also found in \eqref{eq:LMM}, summarizes the linear contribution in
the mixture while the second term models nonlinear interactions
between the materials. The coefficient $\beta_{i,j,p}$ adjusts the
amount of nonlinearities between the components $ \bfm_i$ and
$\bfm_j$ in the $p$th pixel. Several alternatives for imposing
constraints on these nonlinear coefficients have been suggested.
Similarly to \cite{Somers2009}, Nascimento and Dias assume in
\cite{Nascimento2009spie} that the (linear) abundance and
nonlinearity coefficients obey
\begin{equation}
\label{eq:nascimento_model}
  \left\{
     \begin{array}{ll}
       a_{r,p}\geq 0, & \forall p,\ \forall r \\
       \beta_{i,j,p}\geq 0, & \forall p,\ \forall i\neq j\\
       \multicolumn{2}{c}{\sum_{r=1}^R a_{r,p} + \sum_{i=1}^{R-1}\sum_{j=i+1}^R \beta_{i,j,p} = 1.}
     \end{array}
   \right.
\end{equation}
It is worth noting that, from \eqref{eq:nascimento_model}, this
Nascimento model (NM), also used in \cite{Raksuntorn2010}, can be
interpreted as a linear mixing model with additional virtual
endmembers. Indeed, considering $\bfm_i \odot \bfm_j$ as a pure
component spectral signature with corresponding abundance
$\beta_{i,j,p}$, the model in \eqref{eq:nascimento_model} can be
rewritten
\begin{equation*}
  \bfy_p = \sum_{s=1}^{\tilde{R}} \tilde{a}_{s,p} \tilde{\bfm}_s +
  \bfn_p
\end{equation*}
with the positivity and additivity constraints in
\eqref{eq:LMM_constraints} where
\begin{equation*}
  \left\{
     \begin{array}{lll}
       \tilde{a}_{s,p} \triangleq a_{r,p}, & \tilde{\bfm}_s \triangleq \bfm_r           & s=1,\ldots,R \\
       \tilde{a}_{s,p} \triangleq \beta_{i,j,p}, & \tilde{\bfm}_s \triangleq \bfm_i \odot \bfm_j           & s=R+1,\ldots,\tilde{R} \\
     \end{array}
   \right.
\end{equation*}
and $\tilde{R}=\frac{1}{2}R(R+1)$. This NM reduces to the LMM when
$\tilde{a}_{s,p}=0$ for $s=R+1,\ldots,\tilde{R}$.

Conversely, in \cite{Fan2009}, Fan and his co-authors have fixed the
nonlinearity coefficients as functions of the (linear) abundance
coefficients themselves: $\beta_{i,j,p} = a_{i,p}a_{j,p}$ ($i\neq
j$). The resulting model, called the Fan Model (FM) in what follows,
is thus fully described by the mixing equation
\begin{equation}
\label{eq:fan_model}
  \bfy_p = \sum_{r=1}^R a_{r,p}\bfm_r + \sum_{i=1}^{R-1}\sum_{j=i+1}^R a_{i,p}a_{j,p} \bfm_i \odot \bfm_j +
  \bfn_p
\end{equation}
subject to the constraints in \eqref{eq:LMM_constraints}. One
argument to explain the direct relation between the abundances and
the nonlinearity coefficients is the following: if the $i$th
endmember is absent in the $p$th pixel, then $a_{i,p}=0$ and there
are no interactions between $\bfm_i$ and the other materials
$\bfm_j$ ($j\neq i$). More generally, it is quite natural to assume
that the quantity of nonlinear interactions in a given pixel between
two materials is directly related to the quantity of each material
present in that pixel. However, it is clear that this model does not
generalize the LMM, which can be a restrictive property.

More recently, to alleviate this issue, the generalized bilinear
model (GBM) has been proposed in \cite{Halimi2011} by setting
$\beta_{i,j,p}= \gamma_{i,j,p}a_{i,p}a_{j,p}$
\begin{equation}
\label{eq:gbm}
  \bfy_p = \sum_{r=1}^R a_{r,p}\bfm_r + \sum_{i=1}^{R-1}\sum_{j=i+1}^R \gamma_{i,j,p}a_{i,p}a_{j,p} \bfm_i \odot \bfm_j +
  \bfn_p.
\end{equation}
where the interaction coefficient $\gamma_{i,j,p}\in(0,1)$
quantifies the nonlinear interaction between the spectral components
$\bfm_i$ and $\bfm_j$. This model has the same interesting
characteristic as the FM: the amount of nonlinear interactions is
governed by the presence of the endmembers that linearly interact.
In particular, if an endmember is absent in a pixel, there is no
nonlinear interaction supporting this endmember. However, it also
has the significant advantage of generalizing both the LMM when
$\gamma_{i,j,p}=0$ and the FM when $\gamma_{i,j,p}=1$. Having
$\gamma_{i,j,p}>0$ indicate that only constructive interactions are
considered.

\begin{figure*}[b!]
\centering \fbox{
\parbox{0.98\textwidth}{
%\begin{minipage}{1\textwidth}
\textbf{\emph{On the use of geometrical LMM-based EEAs to identify
nonlinearly mixed endmembers \vspace{-0.4cm}}}\hfill \hfil
\linebreak\newline The first automated spectral unmixing algorithms,
proposed in the 1990's, were based on geometrical concepts and were
designed to identify endmembers as pure pixels (see
\cite{Bioucas2012jstars} and \cite{Ma2013} for comprehensive reviews
of geometrical linear unmixing methods). It is worth noting that
this class of algorithms does not explicitly rely on the assumption
of pixels coming from linear mixtures. They only search for
endmembers as extremal points in the hyperspectral dataset. Provided
there are pure pixels in the analyzed image, this might indicate
that some of these geometrical approaches can be still valid for
nonlinear mixtures that preserve this property, such as the GBM and
the FM as illustrated in Fig. \ref{fig:simplexes}. } }
\end{figure*}

\begin{figure}[h!]
  \centering
  \vspace{-0.15cm}
  \includegraphics[width=\figwidth]{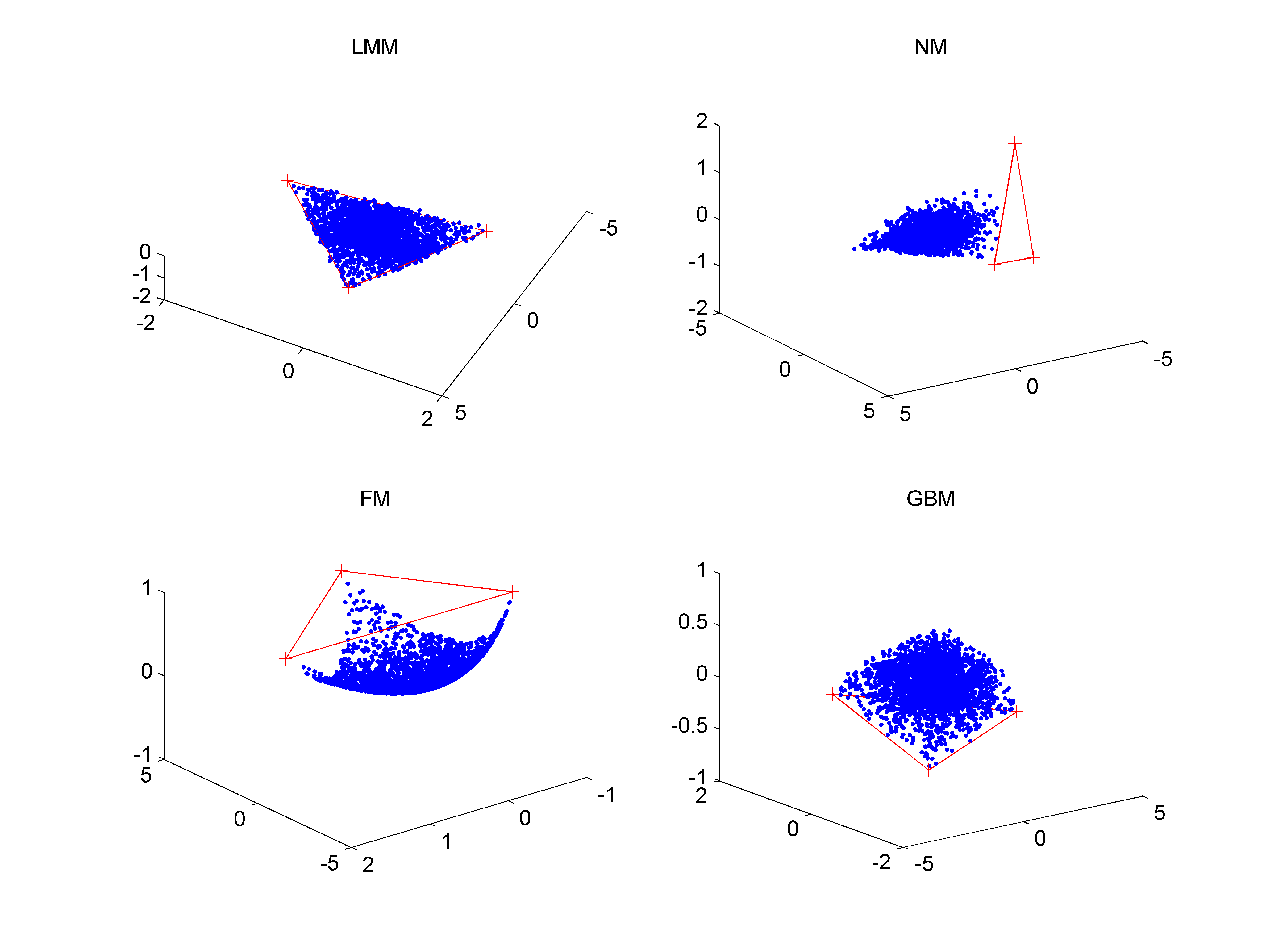}
    \vspace{-0.3cm}
  \caption{Clusters of observations generated according to the LMM, the NM, the
  FM and the GBM (blue) and the corresponding endmembers (red crosses).} \label{fig:simplexes}
  \vspace{-0.15cm}
\end{figure}

For illustration, synthetic mixtures of $R=3$ spectral components
have been randomly generated according to the LMM, NM, FM and GBM.
The resulting data set are represented in the space spanned by the
three principal eigenvectors (associated with the three largest
eigenvalues of the sample covariance matrix of the data) identified
by a principal component analysis in Fig.~\ref{fig:simplexes}. These
plots illustrate an interesting property for the considered dataset:
the spectral signatures of the pure components are still extremal
points, i.e., vertices of the clusters, in the cases of FM and GBM
mixtures contrary to the NM. In other words, geometrical endmember
extraction algorithms (EEAs) and, in particular, those that are
looking for the simplex of largest volume (see \cite{Ma2013} for
details), may still be valid for the FM and the GBM under the
assumption of weak nonlinear interactions.

All these bilinear models only include between-component
interactions $\bfm_i \odot \bfm_j$ with $i\neq j$ but no
within-component interactions $\bfm_i \odot \bfm_i$. Finally, in
\cite{Meganem2013}, the authors derived a nonlinear mixing model
using a RT model applied to a simple canyon-like urban scene.
Successive approximations and simplifying assumptions lead to the
following linear-quadratic mixing model (LQM)
\begin{equation}
\label{eq:meganem}
  \bfy_p = \sum_{r=1}^R a_{r,p}\bfm_r + \sum_{i=1}^{R}\sum_{j=i}^R \beta_{i,j,p} \bfm_i \odot \bfm_j +
  \bfn_p
\end{equation}
with the positivity and additivity constraints in
\eqref{eq:LMM_constraints} and $\beta_{i,j,p} \in (0,1)$. This model
is similar to the general formulation of the bilinear models in
\eqref{eq:bilinear_model}, with the noticeable difference that the
nonlinear contribution includes quadratic terms $\bfm_i \odot \bfm_i
$. This contribution also shows that it is quite legitimate to
include the termwise products $\bfm_i \odot \bfm_j$ as additional
components of the standard linear contribution, which is the core of
the bilinear models described in this section.

\subsection{Other approximating physics-based models}

To describe both macroscopic and microscopic mixtures,
\cite{Close2012spieb} introduces a dual model composed of two terms
\begin{equation*}
  \bfy_p = \sum_{r=1}^R a_{r,p}\bfm_r + a_{R+1,p}\calR\left(\sum_{r=1}^R f_{r,p} \bfw_r\right) +
\bfn_p.
\end{equation*}
The first term is similar to the one encountered in LMM and comes
from the macroscopic mixing process. The second one, considered as
an additional endmember with abundance $a_{R+1,p}$, describes the
intimate mixture by the average single-scattering albedo
\cite{Hapke1981} expressed in the reflective domain by the mapping
$\calR\left(\cdot\right)$.

Altmann \emph{et al.} have proposed in \cite{Altmann2012ip} an
approximating model able to describe a wide class of nonlinearities.
This model is obtained by performing a second-order expansion of the
nonlinearity defining the mixture. More precisely, the $p$th
observed pixel spectrum is defined as a nonlinear transformation
$\bfg_p\left(\cdot\right)$ of a linear mixture of the endmember
spectra
\begin{equation}
\label{eq:ppnmm_model}
  \bfy_p = \bfg_p\left(\sum_{r=1}^R a_{r,p}\bfm_r\right) + \bfn_p
\end{equation}
where the nonlinear function $\bfg_p$ is defined as a second order
polynomial nonlinearity parameterized by the unique nonlinearity
parameter $b_p$
\begin{equation}
\label{eq:nonlinearity_parameter}
     \begin{array}{rcl}
       \bfg_p : &  (0,1)^L & \rightarrow  \mathbb{R}^L\\
              &   \bfx   & \mapsto \left[x_1 + b_p x_1^2,\ldots,x_L + b_p x_L^2 \right]^T
%\left(
%\begin{array}{c}
%    x_1 + b_p x_1^2 \\
%    \ldots \\
%    x_L + b_p x_L^2 \\
%\end{array}
%\right)
\end{array}
\end{equation}
This model can be rewritten
\begin{equation*}
  \bfy_p = \bfM\bfa_p + b_p \left(\bfM\bfa_p\right) \odot
\left(\bfM\bfa_p\right)+ \bfn_p
\end{equation*}
where $\bfM=\left[\bfm_1,\ldots,\bfm_R\right]$ and
$\bfa_p=\left[a_{1,p},\ldots,a_{R,p}\right]^T$. The parameter $b_p$
tunes the amount of nonlinearity present in the $p$th pixel of the
image and this model reduces to the standard LMM when $b_p=0$. It
can be easily shown that this polynomial post-nonlinear model (PPNM)
includes bilinear terms $\bfm_i \odot \bfm_j$ ($i \neq j$) similar
to those defining the FM, NM and GBM, as well as quadratic terms
$\bfm_i \odot \bfm_i$ similar to the LQM in \eqref{eq:meganem}. This
PPNM has been shown to be sufficiently flexible to describe most of
the bilinear models introduced in this section \cite{Altmann2012ip}.

\subsection{Limitations a pixel-wise nonlinear SU}

Having reviewed the above physics-based models, an important remark
must be made. It is important to note that these models do not take
into account spatial interactions from materials present in the
neighborhood of the targeted pixel. It means that these bilinear
models only consider scattering effects in a given pixel induced by
components that are present in this specific pixel. This is a strong
simplifying assumption that allows  the model parameters (abundance
and nonlinear coefficients) to be estimated pixel-by-pixel in the
inversion step. Note however that the problem of taking adjacency
effects  into account, i.e., nonlinear interactions coming from
spectral interference caused by atmospheric scattering, has been
addressed in an unmixing context in \cite{Burazerovic2013jstars}.

%%% ALGORITHMS
\section{Nonlinear unmixing algorithms}
\label{sec:algorithms}

Significant promising approaches have been proposed to nonlinearly
unmix hyperspectral data. A wide class of nonlinear unmixing
algorithms rely explicitly on a nonlinear physics-based parametric
model, as detailed earlier. Others do not require definition of the
mixing model and rely on very mild assumptions regarding the
nonlinearities. For these two classes of approaches, unmixing
algorithms have been considered under two different scenarios,
namely supervised or unsupervised, depending on the available prior
knowledge on the endmembers. When the endmembers are known,
supervised algorithms reduce to estimating the abundance
coefficients in a single supervised inversion step. In this case,
the pure spectral signatures present in the scene must have been
previously identified. For instance, they use prior information or
suboptimal linear EEA. Indeed, as previously noted, when considering
weakly nonlinearly mixed data, the LMM-based EEA may produce good
endmember estimates when there are pure pixels in the dataset. In
contrast, an unsupervised unmixing algorithm jointly estimates the
endmembers and the abundances. Thus the unmixing problem becomes
even more challenging, since a blind source separation problem  must
be solved.

\subsection{Model-based parametric nonlinear unmixing algorithms}

Given a nonlinear parametric model, SU can be formulated as a
constrained nonlinear regression or a nonlinear source separation
problem, depending on whether the endmember spectral signatures are
known or not. When dealing with intimate mixtures, some authors have
proposed converting the measured reflectance into a single
scattering albedo average; since this obeys a linear mixture, the
mass fractions associated with each endmember can be estimated using
a standard linear unmixing algorithm. This is the approach adopted
in \cite{Mustard1989} and \cite{Nascimento2010} for known and
unknown endmembers, respectively. To avoid the functional inversion
of the reflectance measurements into the single scattering albedo, a
common approach is to use neural-networks (NN) to learn this
nonlinear function. This is the strategy followed by Guilfoyle
\emph{et al.} in \cite{Guilfoyle2001}, for which several
improvements have been proposed in \cite{Altmann2011igarssRBF} to
reduce the computationally intensive learning step. In these
NN-based approaches, the endmembers are assumed to be known
a-priori, and are required to train the NN. Other NN-based
algorithms have been studied in
\cite{JPlaza2004,Plaza2005igarss,Plaza2007igarss,Plaza2009pr}.

For the bilinear models introduced previously, supervised nonlinear
optimization methods have been developed based on the assumption
that the endmember matrix $\bfM$ is known. When the observed pixel
spectrum $\bfy_p$ is related to the parameters of interest
$\boldsymbol{\theta}_p$ (a vector containing the abundance
coefficients as well as any other nonlinearity parameters) through
the function $\boldsymbol{\varphi}(\bfM,\cdot)$, unmixing the pixel
$\bfy_p$ consists of solving the following minimization problem
\begin{equation}
    \label{eq:problem.parametric}
  \hat{\boldsymbol{\theta}}_p =
  \operatornamewithlimits{argmin}_{\boldsymbol{\theta}} \left\|\bfy_p-
  \boldsymbol{\varphi}(\bfM,\boldsymbol{\theta}) \right\|_2^2.
\end{equation}
This problem raises two major issues: i) the nonlinearity of the
criterion resulting from the underlying nonlinear model
$\boldsymbol{\varphi}(\cdot)$ and ii) the constraints that have to
be satisfied by the parameter vector $\boldsymbol{\theta}$. Since
the NM can be interpreted as a linear mixing model with additional
virtual endmembers, estimation of the parameters can be conducted
with a linear optimization method as in \cite{Nascimento2009spie}.
In \cite{Fan2009,Halimi2011igarss} dedicated to FM and GBM, the
authors propose to linearize the objective criterion via a
first-order Taylor series expansion of $
\boldsymbol{\varphi}(\cdot)$. Then, the fully constrained least
square (FCLS) algorithm of \cite{Heinz2001} can be used to estimate
parameter vector $\boldsymbol{\theta}$. An alternative algorithmic
scheme proposed in \cite{Halimi2011igarss} consists of resorting to
a gradient descent method where the step-size parameter is adjusted
by a constrained line search procedure enforcing the constraints
inherent to the mixing model. Another strategy initially introduced
in \cite{Halimi2011} for the GBM is based on Monte Carlo
approximations, developed in a fully Bayesian statistical framework.
The Bayesian setting has the great advantage of providing a
convenient way to include the parameter constraints within the
estimation problem, by defining appropriate priors for the
parameters. This strategy has been also considered to unmix the PPNM
\cite{Altmann2012ip}.

When the spectral signatures $\bfM$ involved in these bilinear
models need also to be identified in addition to the abundances and
nonlinearity parameters, more ambitious unmixing algorithms need to
be designed. In \cite{Gader2012whispers}, the authors differentiate
the NM to implement updating rules that generalize the SPICE
algorithm introduced in \cite{Zare2007} for the linear mixing model.
Conversely, NMF-based iterative algorithms have been advocated in
\cite{Yokoya2012igarss} for the GBM defined in \eqref{eq:gbm}, and
in \cite{Meganem2013} for the LQM described in \eqref{eq:meganem}.
More recently, an unsupervised version of the Bayesian PPNM-based
unmixing algorithm initially introduced in \cite{Altmann2012ip} has
been investigated in \cite{Altmann2013sub}.

Adopting a geometrical point-of-view, Heylen and Scheunders propose
in \cite{Heylen2012grsl} an integral formulation to compute geodesic
distances on the nonlinear manifold induced by the GBM. The
underlying idea is to derive an EEA that identifies the simplex of
maximum volume contained in the manifold defined by the GBM-mixed
pixels.

\subsection{Model-free nonlinear unmixing algorithms}

When the nonlinearity defining the mixing is unknown, the SU problem
becomes even more challenging. In such cases, when the endmember
matrix $\bfM$ is fixed a priori, a classification approach can be
adopted to estimate the abundance coefficients  which can be solved
using support vector machines \cite{Plaza2004spie,Li2005igarss}.
Conversely, when the endmember signatures are not known, a
geometrical-based unmixing technique can be used, based on
graph-based approximate geodesic distances~\cite{Heylen2011jstsp},
or manifold learning techniques
\cite{Nguyen2012igarss,Licciardi2012igarss}. Another promising
approach is to use nonparametric methods based on reproducing
kernels
\cite{Broadwater2007igarss,Broadwater2009whispers,Chen2013sp,Chen2013tgrs,Li2012sivp,Nguyen2013EAS}
or on Gaussian processes \cite{Altmann2013sp} to approximate the
unknown nonlinearity. These two later techniques are described
below.

Nonlinear algorithms operating in reproducing kernel Hilbert spaces
(RKHS) have received considerable interest in the machine learning
community, and have proved their efficiency in solving nonlinear
problems. Kernel-based methods have been widely considered for
detection and classification in hyperspectral images. Surprisingly,
nonlinear unmixing approaches operating in RKHS have been
investigated in a less in-depth way. The algorithms derived in
\cite{Broadwater2007igarss,Broadwater2009whispers} were mainly
obtained by replacing each inner product between endmember spectra
in the cost functions to be optimized by a kernel function. This can
be viewed as a nonlinear distortion map applied to the spectral
signature of each material, independently of  their interactions.
This principle can be extremely efficient in solving detection and
classification problems as a proper distortion can increase the
detectability or separability of some patterns. It is however of
little physical interest in solving the unmixing problem because the
nonlinear nature of the mixtures is not only governed by individual
spectral distortions, but also by nonlinear interactions between the
materials. In \cite{Chen2013sp}, a new kernel-based paradigm was
proposed to take the nonlinear interactions of the endmembers into
account, when these endmembers are assumed to be a priori known. It
solves the optimization problem

\begin{equation}
      \label{eq:problem.functional}
     \mathop{\min}_{\psi_{\boldsymbol{\theta}}\in\calH}
      \sum_{\ell=1}^L \left[y_{\ell,p}-\psi_{\boldsymbol{\theta}}\!\left(\bfm_{\lambda_\ell}\right)\right]^2 + \mu\|\psi_{\boldsymbol{\theta}}\|_\calH^2
\end{equation}
where $\bfm_{\lambda_\ell}$ is the vector of the endmember
signatures at the $\ell$-th frequency band, namely,
$\bfm_{\lambda_\ell} = \left[m_{\ell,1},\ldots,m_{\ell,R}\right]^T$,
with $\calH$ a given functional space, and $\mu$ a positive
parameter that controls the trade-off between regularity of the
function $\psi_{\boldsymbol{\theta}}(\cdot)$ and fitting. Again,
$\boldsymbol{\theta}$ is a vector containing the abundance
coefficients as well as any other nonlinearity parameters. It is
interesting to note that \eqref{eq:problem.functional} is the
functional counterpart to \eqref{eq:problem.parametric}, where
$\psi_{\boldsymbol{\theta}}(\cdot)$ defines the nonlinear
interactions between the endmembers assumed to be known in
\cite{Chen2013sp}. Clearly, this strategy may fail if the functional
space $\calH$ is not chosen appropriately. A successful strategy is
to define $\calH$ as an RKHS in order to exploit the so-called
kernel trick. Let $\kappa(\cdot\,,\cdot)$ be the reproducing kernel
of $\calH$. The RKHS $\calH$ must be carefully selected via its
kernel in order to make it flexible enough to capture wide classes
of nonlinear relationships, and to reliably interpret a variety of
experimental measurements. In order to extract the mixing ratios of
the endmembers, the authors in \cite{Chen2013sp} focus their
attention on partially linear models, resulting in the so-called
K-HYPE SU algorithm. More precisely, the function
$\psi_{\boldsymbol{\theta}}(\cdot)$ in problem
\eqref{eq:problem.functional} is defined by an LMM parameterized by
the abundance vector $\bfa$, combined with a nonparametric term,
\begin{equation}
    \label{eq:map}
            \psi_{\boldsymbol{\theta}}(\bfm_{\lambda_\ell})   = \bfa^\top\bfm_{\lambda_\ell}+\psi_\text{nlin}(\bfm_{\lambda_\ell})
\end{equation}
possibly subject to the constraints in \eqref{eq:LMM_constraints},
where $\psi_\text{nlin}$ can be any real-valued function of an RKHS
denoted by $\calH_\text{nlin}$. This model generalizes the standard
LMM, and mimics the PPNM when $\calH_\text{nlin}$ is defined to be
the space of polynomial functions of degree $2$. Remember that the
latter is induced by the polynomial kernel
$\kappa(\bfm_{\lambda_\ell},\bfm_{\lambda_{\ell'}})=(\bfm_{\lambda_\ell}^\top\bfm_{\lambda_{\ell'}})^q$
of degree $q=2$. More complex interaction mechanisms can be
considered by simply changing
$\kappa(\bfm_{\lambda_\ell},\bfm_{\lambda_{\ell'}})$.  By virtue of
the reproducing kernel machinery,  the problem can still be solved
in the framework of \eqref{eq:problem.functional}.

Another strategy introduced in \cite{Altmann2013sp} considers a
kernel-based approach for unsupervised nonlinear SU based on a
nonlinear dimensionality reduction using a Gaussian process latent
variable model (GPLVM). In this work, the authors have used a
particular form of kernel which extends the generalized bilinear
model in \eqref{eq:gbm}. The algorithm proposed in
\cite{Altmann2013sp} is unsupervised in the sense that the
endmembers contained in the image and the mixing model are not
known. Only the number of endmembers is assumed to be known. As a
consequence, the parameters to be estimated are the kernel
parameters, the endmember spectra and the abundances for all image
pixels. The main advantage of GPLVMs is their capacity to accurately
model many different nonlinearities. GPLVMs construct a smooth
mapping from the space of fractional abundances to the space of
observed mixed pixels that preserves dissimilarities. This strategy
has been also considered in \cite{Nguyen2013EAS} by Nguyen \emph{et
al.}, who solve the so-called pre-image problem
\cite{Honeine2011sigmag} studied in the machine learning community.
In the SU context, it means that pixels that are spectrally
different have different latent variables and thus different
abundance vectors. However, preserving local distances is also
interesting: spectrally close pixels are expected to have similar
abundance vectors and thus similar latent variables. Several
approaches have been proposed to preserve similarities, including
back-constraints and locally linear embedding.

\begin{table*}
\renewcommand{\arraystretch}{1.05}
%\begin{footnotesize}
\begin{center}
\begin{tabular}{|c|c|c|c|c|c|c||c|c|c|c|}
\cline{4-11}
\multicolumn{3}{c|}{}  &  \multicolumn{4}{|c||}{Mixing models -- with pure pixels}  &  \multicolumn{4}{|c|}{Mixing models -- w/o    pure pixels} \\
 \cline{4-11}
\multicolumn{3}{c|}{}  &  LMM  & PPNM&  GBM  & FM &  LMM  & PPNM&  GBM  & FM\\
\hline \multirow{6}*{\rotatebox{90}{Model-based algo.}} &
\multirow{2}*{\rotatebox{0}{LMM}}  & N-FINDR + FCLS
            &   $1.42$   &   $14.1$    &   $7.71$      &  $13.4$
            &$3.78$ & $13.2$ & $6.83$ & $9.53 $\\
 \cline{3-11}
&   &  unsupervised MCMC
            &   $0.64$   &   $12.4$    &   $5.71$      &  $8.14$
            & $0.66 $& $10.9$ & $4.21$ & $3.92$\\
 \cline{2-11}
&  \multirow{2}*{\rotatebox{0}{PPNM}}  & Geodesic + GBA
            &   $1.52$   &   $10.3$    &   $6.04$      &  $12.1$
            & $4.18$& $6.04$ & $4.13$ & $3.74$\\
\cline{3-11} &   &  unsupervised MCMC
            &   $0.39$   &   $0.73$    &   $1.32$      &  $2.14$
            & $0.37$ & $0.81$ & $1.38$ & $2.25$\\
\cline{2-11}
 & \rotatebox{0}{GBM} & Geodesic + GBA
            &   $2.78$   &   $14.3$    &   $6.01$      &  $13.0$
            & $4.18$& $11.1$& $5.02$ & $1.45 $ \\
\cline{2-11} & \rotatebox{0}{FM} & Geodesic + GBA
            &   $13.4$   &   $21.8$    &   $9.90$      &  $3.40$
            & $12.2$& $ 18.1$& $7.17$ & $4.97 $ \\
\hline \multicolumn{3}{|c|}{Geodesic + K-HYPE}
            &   $2.43$   &   $9.71$    &   $5.23$      &  $11.3$
            & $ 2.44$& $5.92$& $3.18$ & $2.58 $\\
\hline
\end{tabular}
\vspace{-0.3cm} \caption{Abundance RNMSEs ($\times 10^{-2}$) for
various linear/nonlinear unmixing scenarios.\label{tab:RMSE_synth}}
\end{center}
\vspace{-0.5cm}
%\end{footnotesize}
\end{table*}

For illustration, a small set of experiments has been conducted to
evaluate some of the model-based and model-free algorithms
introduced above. First, $4$ synthetic images of size $50\times 50$
have been generated by mixing $R=3$ endmember spectra (i.e., green
grass, olive green paint and galvanized steel metal) extracted from
the spectral libraries provided with the ENVI software
\cite{ENVImanual2003}. These $4$ images have been generated
according to the standard LMM \eqref{eq:LMM}, the GBM
\eqref{eq:gbm}, the FM \eqref{eq:fan_model} and the PPNM
\eqref{eq:ppnmm_model}, respectively. For each image, the abundance
coefficient vectors
$\bfa_{p}\triangleq\left[a_{1,p},\ldots,a_{3,p}\right]$
($p=1,\ldots,2500$) have been randomly and uniformly generated in
the admissible set defined by the constraints
\eqref{eq:LMM_constraints}. We have also considered the more
challenging scenario defined by the assumption that there is no pure
pixel (by imposing $a_{r,p}<0.9$, $\forall r,\forall p$). The
nonlinearity coefficients are uniformly drawn in the set $[0, 1]$
for the GBM. The PPNM-parameters $b_p$, $p = 1\ldots,P$ have been
generated uniformly in the set $[-0.3, 0.3]$. For both scenario
(i.e., with of without pure pixels), all images have been corrupted
by an additive independent and identically distributed (i.i.d)
Gaussian noise of variance $\sigma^2=10^{-4}$, which corresponds to
an average signal-to-noise ratio of $20$dB (note that the usual SNR
for most of the spectro-imagers are not below $30$dB). Various
unmixing strategies have been implemented to recover the endmember
signatures and then estimate the abundance coefficients. For
supervised unmixing, the N-FINDR algorithm \cite{Winter1999}  and
its nonlinear geodesic-based counterpart \cite{Heylen2011jstsp} have
been used to extract the endmembers from linear and nonlinear
mixtures, respectively. Then, dedicated model-based strategies were
used to recover the abundance fractions. The fully constrained least
square (FCLS) algorithm \cite{Heinz2001} was used for linear
mixtures. Gradient-based algorithms (GBA) were used for nonlinear
mixtures. The GBA are detailed in \cite{Altmann2013ip},
\cite{Halimi2011igarss} and \cite{Fan2009} for the PPNM, GBM and FM,
respectively. For comparison with supervised unmixing, and to
evaluate the impact of having no pure pixels in these images, joint
estimations of endmembers and abundances was implemented using the
Markov chain Monte Carlo techniques detailed in \cite{Dobigeon2009}
and \cite{Altmann2013sub} for the LMM and PPNM images, respectively.
Finally, the model-free supervised K-HYPE algorithm detailed in
\cite{Chen2013sp} was also coupled with the nonlinear EEA in
\cite{Heylen2011jstsp}. The performance of these unmixing strategies
has been evaluated in term of abundance estimation error measured by
\begin{equation*}
  \textrm{RNMSE} = \sqrt{\frac{1}{RP}\sum_{n=1}^N \left\|\bfa_p - \hat{\bfa}_p\right\|^2}
\end{equation*}
where $\bfa_p$ is the $n$th actual abundance vector and
$\hat{\bfa}_p$ its corresponding estimate. The results are reported
in Table~\ref{tab:RMSE_synth}. These results clearly show that the
prior knowledge of the actual mixing model underlying the
observations is a clear advantage for abundance estimation. However,
in the absence of such knowledge, using an inappropriate model-based
algorithm may lead to poor unmixing results. In such cases, as
advocated before, PPNM seems to be sufficiently flexible to provide
reasonable estimates, whatever the mixing model may be. Otherwise,
one may prefer to resort to model-free based strategy such as
K-HYPE.

%%% DETECTION
\section{Detecting nonlinear mixtures}
\label{sec:detection}

Consideration of nonlinear effects in hyperspectral images can
provide more accurate results in terms of endmember and abundance
identification.  However, working with nonlinear models generally
requires a higher computational complexity than approaches based on
the LMM.  Thus, unmixing linearly mixed pixels using nonlinear
models should be avoided.  Consequently, it is of interest to devise
techniques to detect nonlinearities in the mixing process before
applying any unmixing method.  Linearly mixed pixels can then be
unmixed using linear unmixing techniques, leaving the application of
more involved nonlinear unmixing methods to situations where they
are really necessary. This section describes approaches that have
been recently proposed to detect nonlinear mixing in hyperspectral
images.

\subsection{Detection using a polynomial post-nonlinear model (PPNM)}
\label{subsec:PNMM-detection} 
\begin{figure}[h!]
  \centering
  \vspace{-0.15cm}
  \includegraphics[width=\figwidth]{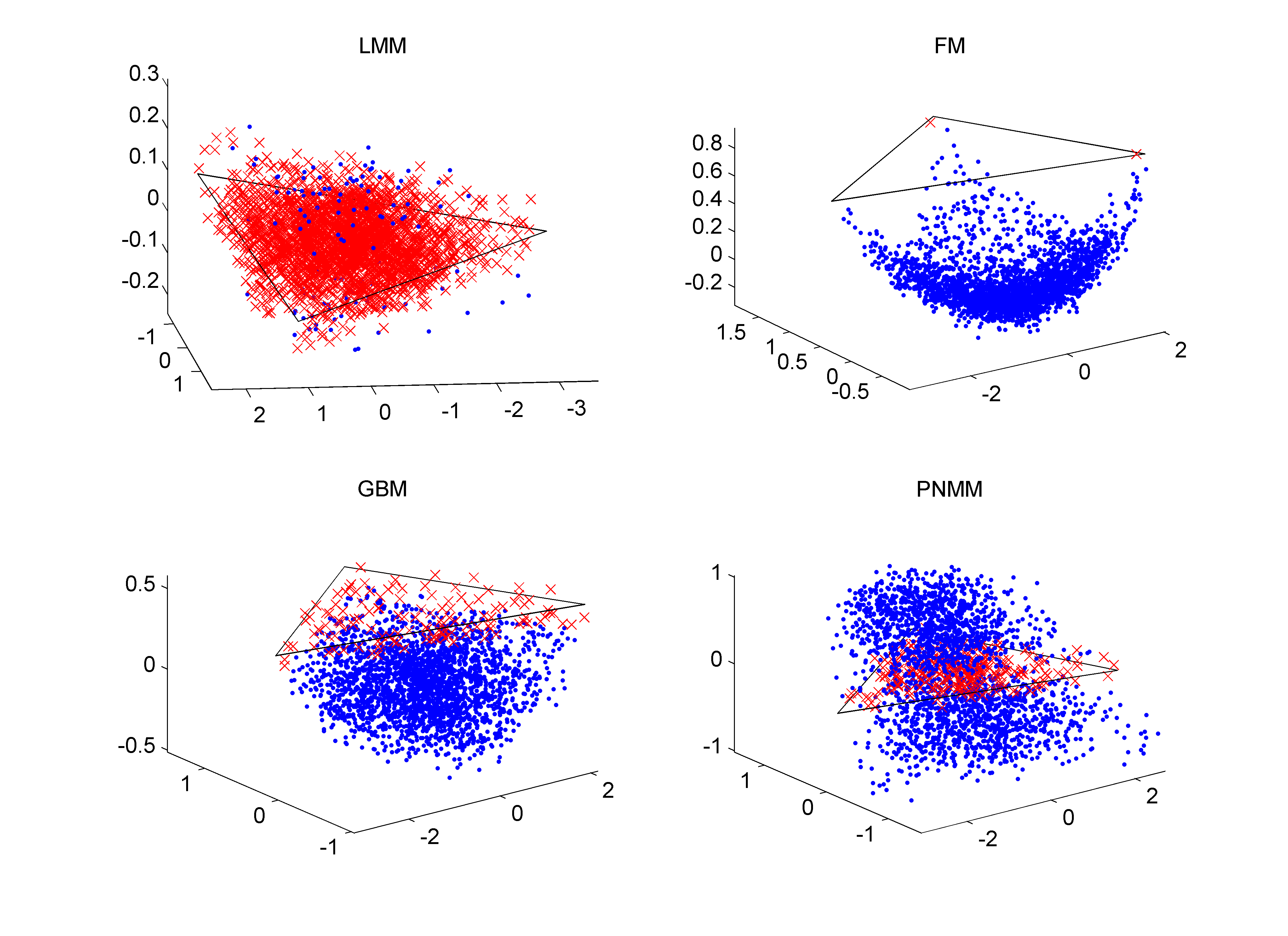}
    \vspace{-0.3cm}
  \caption{Pixels detected as linear (red crosses) and nonlinear (blue dotted)
for the four subimages generated according the LMM, FM, GBM, and
PPNM. Black lines depict the simplex corresponding to the noise-free
case LMM.} \label{fig:detection_new}
  \vspace{-0.15cm}
\end{figure}

One interesting approach for
nonlinearity detection is to assume a parametric nonlinear mixing
model that can model different nonlinearities between the endmembers
and the observations. A model that has been successfully applied to
this end is the PPNM \eqref{eq:ppnmm_model} studied in
\cite{Altmann2012ip,Altmann2013ip}. PPNM assumes the post-nonlinear
mixing described in \eqref{eq:ppnmm_model} with the polynomial
nonlinearity $\bfg_p$ defined in \eqref{eq:nonlinearity_parameter}.
Hence, the nonlinearity is characterized by the parameter $b_p$ for
each pixel in the scene. This parameter can be estimated in
conjunction with the abundance vector $\bfa_p$ and the noise
variance $\sigma^2$. Denote as $s^2(\bfa_p,b_p,\sigma^2)$ the
variance of the  maximum likelihood estimator $\hat{b}_p$ of $b$.
Using the properties of the maximum likelihood estimator, it makes
sense to approximate the distribution of $\hat{b}_p$ by the
following Gaussian distribution
\[
\hat{b}_p \sim \mathcal{N}\left(b_p,s^2(a_p,b_p,\sigma^2)\right).
\]
The nonlinearity detection problem can be formulated as the
following binary hypothesis testing problem
\begin{eqnarray}
\label{eq:detection_pb0} \left\{
    \begin{array}{ll}
        \calH_0  &: \bfy_p \textrm{ is distributed according to the LMM \eqref{eq:LMM}}\\
        \calH_1  &: \bfy_p \textrm{ is distributed according to the
        PPNM \eqref{eq:ppnmm_model}}
    \end{array}
\right.
\end{eqnarray}
Hypothesis $\calH_0$ is characterized by $b_p=0$ whereas nonlinear
models ($\calH_1$) correspond to $b_p\neq0$. Then,
\eqref{eq:detection_pb0} can be rewritten as
\begin{eqnarray}
\label{eq:detection_pb} \left\{
    \begin{array}{lll}
        \calH_0 & : & \hat{b}_p \sim  \mathcal{N}(0,s_{0}^2) \\
        \calH_1 & : & \hat{b}_p \sim  \mathcal{N}(b_p,s_{1}^2) %\quad b\neq0
    \end{array}
\right.
\end{eqnarray}
where $s_0^2=s^2(\bfa_p,0,\sigma^2)$ and
$s_1^2=s^2(\bfa_p,b_p,\sigma^2)$ with $b_p\neq0$. Detection can be
performed using the generalized likelihood ratio test. %defined as
%\begin{eqnarray}
%\label{eq:test} T= \dfrac{\hat{b}_p^2}{s_{0}^2}
%\overset{\calH_1}{\underset{\calH_0}{\gtrless}} \eta.
%\end{eqnarray}
This test accepts $\calH_1$ (resp. $\calH_0$) if the ratio
$T\triangleq \hat{b}_p^2 \slash s_{0}^2$ is greater (resp. lower)
than a threshold $\eta$. As shown in \cite{Altmann2013ip}, the
statistic $T$ is approximately normally distributed  under the two
hypotheses. Consequently, the threshold $\eta$ can be explicitly
related to the probability of false alarm (PFA) and the probability
of detection (PD), i.e., the power of the test. However, this
detection strategy assumes the prior knowledge of the variances
$s_0^2$ and $s_1^2$. In practical applications, Altmann \emph{et
al.} show that have proposed to modify the previous test strategy as
follows \cite{Altmann2013ip}
\begin{eqnarray}
\label{eq:test_star} \hat{T}= \dfrac{\hat{b}_p^2}{\hat{s}_{0}^2}
\overset{\calH_1}{\underset{\calH_0}{\gtrless}} \eta^*
\end{eqnarray}
where $\hat{s}_{0}^2$ can be calculated as
\begin{equation}
\label{eq:CCRLB}
\hat{s}_{0}^2=\mathrm{CCRLB}(0;\hat{\bfa}_p,\hat{\sigma}^2)
\end{equation}
In \eqref{eq:CCRLB}, $\mathrm{CCRLB}$ is the constrained
Cram\'er-Rao lower-bound \cite{Gorman1990} on estimates of the
parameter vector ${\boldsymbol{\theta}}=[\bfa_p^T,b_p,\sigma^2]^T$
under $\calH_0$, and $\left(\hat{\bfa}_p,\hat{\sigma}^2\right)$ is
the MLE of $\left(\bfa_p,\sigma^2\right)$. The performance of the
resulting test is illustrated in Fig. \ref{fig:detection_new} which
shows the pixels detected as linear (red crosses) and nonlinear
(blue dots) when generated according to various mixing models (LMM,
FM, GBM and PPNM).

\subsection{Robust model-free detection}

The detector discussed in the previous section assumes a specific
nonlinear mixing model under the alternative hypothesis.  However,
there are situations where the actual mixing does not obey any
available model. It is also possible that there is insufficient
information to opt for any existing nonlinearity model. In these
cases, it is interesting to address the problem of determining
whether an observed pixel is a linear function of endmembers or
results from a generic nonlinear mixing.

One may consider the LMM \eqref{eq:LMM} and the
hyperplane $\mathcal{P}$ defined by
\begin{equation}
\label{eq:plan} \mathcal{P} : \left \lbrace \boldsymbol{z}_p \bigg|
\boldsymbol{z}_p = {\mathbf M}\bfa_p , \sum_{r=1}^{R}{a_{r,p}}=1
\right \rbrace.
\end{equation}
In the noise-free case, the hyperplane $\mathcal{P}$ lies in an
$(R-1)$-dimensional subspace embedding all observations distributed
according to the LMM.  On the other hand, consider the general
nonlinear mixing model
\begin{equation}
\label{eq:NLMM} \bfy_p = {\mathbf M}\bfa_p + \boldsymbol{\mu}_p +
\bfn_p
\end{equation}
where $\boldsymbol{\mu}_p$ is an $L \times 1$ deterministic vector
that does not belong to $\mathcal{P}$, i.e., $\boldsymbol{\mu}_p
\notin \mathcal{P}$ and $\bfa_p$ satisfies the constraints
\eqref{eq:LMM_constraints}. Note that a similar nonlinear mixing
model coupled with a group-sparse constraint on $\boldsymbol{\mu}_p$
has been explicitly adopted in
\cite{Altmann2013bsub,Dobigeon2013whispers} to make more robust the
unmixing of hyperspectral pixels. In \eqref{eq:NLMM},
$\boldsymbol{\mu}_p$ can be a nonlinear function of the endmember
matrix ${\mathbf M}$ and/or the abundance vector $\bfa_p$ and should
be denoted as $\boldsymbol{\mu}_p({\mathbf M},\bfa_p)$
\cite{Altmann2013icassp}. However, the arguments ${\mathbf M}$ and
$\bfa_p$ are omitted here for brevity. Given an observation vector
$\bfy_p$, the detection of nonlinear mixtures can be formulated as
the following binary hypothesis testing problem
\begin{eqnarray*} %\label{eq:detection_pb0}
\left\{
    \begin{array}{lll}
        \calH_0 & : & \bfy_p \textrm{ is distributed according to the LMM \eqref{eq:LMM}}\\
        \calH_1 & : & \bfy_p \textrm{ is distributed according to the model \eqref{eq:NLMM}}.
    \end{array}
\right.
\end{eqnarray*}
Using the statistical properties of the noise $\bfn_p$, we obtain
$\mathrm{E}[\bfy_p|\calH_0]={\mathbf M} \bfa_p \in \mathcal{P}$
whereas $\mathrm{E}[\bfy_p|\calH_1]={\mathbf M} \bfa_p +
\boldsymbol{\mu}_p \notin \mathcal{P}$. As a consequence, it makes
sense to consider the squared Euclidean distance
\begin{equation}
\label{eq:delta} \delta^2(\bfy_p) = \underset{\boldsymbol{z}_p \in
\mathcal{P}}{\min} \|{\bfy_p - \boldsymbol{z}_p}\|^2
\end{equation}
between the observed pixel $\bfy_p$ and the hyperplane $\mathcal{P}$
to decide which hypothesis ($\calH_0$ or $\calH_1$) is true.

As shown in \cite{Altmann2013icassp}, the test statistic
$\delta^2(\bfy_p)$ is distributed according to $\chi^2$ distribution
under the two hypotheses $\calH_0$ and $\calH_1$. The parameters of
this distribution depend on the known matrix ${\bf M}$, the noise
variance $\sigma^2$ and the nonlinearity vector
$\boldsymbol{\mu}_p$. If $\sigma^2$ is known, the distribution of
$\delta^2(\bfy_p)$ is perfectly known under $\calH_0$ and partially
known under $\calH_1$. In this case, one may employ a statistical
test that does not depend on $\boldsymbol{\mu}_p$. This test accepts
$\calH_1$ (resp. $\calH_0$) if the ratio $T \triangleq
\delta^2(\bfy_p)\slash \sigma^2$ is greater (resp. lower) than a
threshold $\eta$. As in the PPNM-based detection procedure, the
threshold $\eta$ can be related to the PFA and PD through
closed-form expressions. In particular, it is interesting to note
that the PD is intrinsically related to a non-Euclidean norm of the
residual component $\boldsymbol{\mu}_p$ (see \cite[Eq.
(11)]{Altmann2013icassp}), which is unfortunately unknown in most
practical applications. %
%such as
%\begin{equation}
%\label{eq:test_0} T = \dfrac{\delta^2(\bfy_p)}{\sigma^2}
%\overset{\calH_1}{\underset{\calH_0}{\gtrless}} \eta.
%\end{equation}
%\begin{eqnarray}
%\label{eq:PFA}
%P_\mathrm{FA} & = & \mathbb{P}\left[T> \eta \bigg| H_0\right]
%\end{eqnarray}
%or, equivalently, $\eta = \mathrm{F}_{\chi^2_K}^{-1}\left(1 - P_\mathrm{FA}\right)$ where $\mathrm{F}_{\chi^2_K}^{-1}$ is the inverse cumulative
%distribution function of the $\chi^2_K$-distribution. For a
%given $\boldsymbol{\mu}$, the power $P_\mathrm{D}(\boldsymbol{\mu})$ of the test is
%\begin{eqnarray}
%P_\mathrm{D}(\boldsymbol{\mu})= 1- \mathrm{F}_{\chi_K^2(\lambda)}(\eta)
%\end{eqnarray}
%where $\mathrm{F}_{\chi_K^2(\lambda)}$ is the cumulative
%distribution function of the $\chi_K^2(\lambda)$-distribution
%and $\lambda=\sigma^{-2}\boldsymbol{\mu}^T{\bf H}\boldsymbol{\mu}$. Notice that the
%probability of detection (PD) $P_\mathrm{D}(\boldsymbol{\mu})$ is an increasing
%function of $\lambda$ for a fixed threshold $\eta$. This makes
%sense, as the higher the power $\boldsymbol{\mu}^T{\bf H}\boldsymbol{\mu}$ of the
%nonlinearity orthogonal to $\mathcal{H}$, the better the detection
%performance. Moreover, the lower the noise variance, the better the
%nonlinearity detection.
If the noise variance $\sigma^2$ is unknown, which is the case in
most practical applications, one can replace $\sigma^2$ with an
estimate $\hat{\sigma}^2$, leading to
\begin{equation}
\label{eq:test_1} T^* \triangleq
\dfrac{\delta^2(\bfy_p)}{\hat{\sigma}^2}
\overset{\calH_1}{\underset{\calH_0}{\gtrless}} \eta
\end{equation}
where $\eta$ is the threshold computed as previously indicated. The
PFA and PD of the test \eqref{eq:test_1} are then explicitly
obtained using cumulative distribution functions of the $\chi^2$
distribution.
%given by
%\begin{eqnarray}
%\label{eq:PFA_PD_test1}
%P_\mathrm{FA}^* & = & \mathbb{P}\left[T^*> \eta \bigg| H_0\right]  = \mathbb{P}\left[T> \dfrac{\hat{\sigma}^2}{\sigma^2}\eta \bigg| H_0\right]\nonumber\\
%P_\mathrm{D}^*(\boldsymbol{\mu}) & = & \mathbb{P}\left[T^*  > \eta\bigg|H_1\right] = \mathbb{P}\left[T > \dfrac{\hat{\sigma}^2}{\sigma^2}\eta\bigg|H_1\right].
%\end{eqnarray}
It was shown in \cite{Altmann2013icassp} that the better the
estimation of $\sigma^2$, the closer the distributions of $T$ and
$T^*$ and thus the closer the performances of the two corresponding
tests. Several techniques can be used to estimate $\sigma^2$. For
instance, $\hat{\sigma}^2$ has been estimated in
\cite{Altmann2013icassp} through an eigen-analysis of the sample
covariance matrix of a set of pixels assumed to share the same
variance. The value of $\hat{\sigma}^2$ was determined as the
average of the smallest eigenvalues of the sample covariance matrix.
The accuracy of the estimator is a function of the number of
eigenvalues considered. It was shown in \cite{Altmann2013icassp}
that a PFA smaller (resp. larger) than $P_\mathrm{FA}^*$ is obtained
if $\hat{\sigma}^2 > \sigma^2$ (resp. $\hat{\sigma}^2 < \sigma^2$).

%%% CONCLUSION
\section{Conclusions and open challenges}

To overcome the intrinsic limitations of the linear mixing model,
several recent contributions have been made for modeling of the
physical processes that underly hyperspectral observations. Some
models attempt to account for between-material interactions
affecting photons before they reach the spectro-imager. Based on
these models, several parametric algorithms have been proposed to
solve the resulting nonlinear unmixing problem. Another class of
unmixing algorithms attempts to avoid the use of any rigid nonlinear
model by using nonparametric machine learning-inspired techniques.
The price to pay for handling nonlinear interactions induced by
multiple scattering effects or intimate mixtures is the
computational complexity and a possible degradation of unmixing
performance when processing large hyperspectral images. To overcome
these difficulties, one possible strategy consists of detecting
pixels subjected to nonlinear mixtures in a pre-processing step. The
pixels detected as linearly mixed can then benefit from the huge and
reliable literature dedicated to the linear unmixing problem. The
remaining pixels (detected as nonlinear) can then be the subject of
particular attention. %Finally, this paper demonstrates that the
%nonlinear unmixing problem represents an exciting issue that
%requires the expertise from signal and image processing researchers
%with various
%methodological backgrounds.\\

This paper has described developments methods in nonlinear mixing
for hyperspectral imaging. Several important several interesting
challenges remain. First of all, better integration of algorithmic
approaches and physical models have the potential to greatly improve
non-linear unmixing performance. By fully accounting for complex RT
effects, such as scattering, dispersion, and beam interaction depth,
a physical model can guide the choice of simplified mathematical and
statistical models. Preliminary results have been recently
communicated in \cite{Tits2012igarss}, based on in situ measurements
coupled with simulation tools (e.g., ray-tracing techniques). A
second challenge is to develop unmixing models that take
heterogeneity of the medium into account. Heterogeneous regions
consist of combinations of linear, weakly non-linear and strongly
non-linear pixels. The detection strategies detailed above might be
one solution to tackle this problem since they are able to locate
the areas where a non-linear model may outperform a linear model and
vice-versa. Another approach adopted in
\cite{Altmann2013bsub,Dobigeon2013whispers}, which works well when
there are only a few non-linear subregions, consists of using a
statistical outlier approach to identify the non-linear pixels.
Moreover, as any nonlinear blind source separation problem, deriving
flexible unsupervised unmixing algorithms is still a major
challenge, especially if one wants to go one step further than a
crude pixel-by-pixel analysis by exploiting spatial information
inherent to these images. Finally, we observe that the presence of
nonlinearity in the observed spectra is closely related to the
number $R$ of endmembers which is usually unknown. For example, in
analogy to kernelization in machine learning, after non-linear
transformation, a nonlinear mixture of $R$ components can often be
represented  as a linear mixture of $\tilde{R}$ endmembers, with
$\tilde{R}>R$. Recent advances in manifold learning and
dimensionality estimation are promising approaches to the non-linear
unmixing problem.

%%% ACKNOWLEDGMENTS

\section*{Acknowledgments}
Part of this work has been funded by the Hypanema ANR Project
n$^\circ$ANR-12-BS03-003 and the MADONNA project supported by INP
Toulouse, France. Some results obtained in this paper result from
fruitful discussions during the ``CIMI Workshop on Optimization and
Statistics in Image Processing'' (Toulouse, June 24-28, 2013). The
authors are grateful to Yoann Altmann (University of Toulouse) for
sharing his numerous MATLAB codes and for stimulating discussions.

%%% BIBLIOGRAPHY

\newpage
%\footnotesize
\bibliographystyle{ieeetran}
\bibliography{strings_all_ref,biblio_hyperspectral}

\end{document}